# Modulation of charge density waves in a twirling moiré superlattice

Qian Fang[1,2#], Yanhao Shi[1,2#], Jingyi Duan[3#], Hui Guo[1,2#]*, Yikai Chen[3], Senhao Lv[1,2], Jiayi Wang[1,2], Zhongyi Cao[1,2], Jiayi Huang[1,2], Siyu Xu[1,2], Haitao Yang[1,2], Wei Jiang[3,4]*, Hui Chen[1,2]*, and Hong-Jun Gao[1,2]

[1] Beijing National Center for Condensed Matter Physics and Institute of Physics, Chinese Academy of Sciences, Beijing 100190, China

[2] School of Physical Sciences, University of Chinese Academy of Sciences, Beijing 100190, China

[3] Key Laboratory of advanced Optoelectronic Quantum Architecture and Measurement (MOE), School of Physics, Beijing Institute of Technology, Beijing 100081, China

[4] International Center for Quantum Materials, Beijing Institute of Technology, Zhuhai 519000, China

[#]These authors contributed equally to this work

*Correspondence to: hchenn04@iphy.ac.cn, guohui@iphy.ac.cn, wjiang@bit.edu.cn

## Abstract

Twisted moiré superlattices provide a powerful platform for engineering correlated electronic states, yet continuous nanoscale manipulation of collective orders remains largely unexplored. Here, we report the modulation of charge density wave (CDW) states in a twisted twirling moiré superlattice formed by monolayer $VTe_2$ on superconducting $NbSe_2$. Scanning tunneling microscopy/spectroscopy reveals that the intrinsic long-range CDW of monolayer $VTe_2$ is reconstructed into inequivalent local phases with distinct stability and coherence within a single moiré unit cell, including suppressed CDW order and enhanced short-range CDW correlations persisting to room temperature. First-principles calculations attribute the reconstructed CDW landscape to strong local strain variation, with compressive strain significantly stabilizing the charge order. The reconstructed CDW further exhibits an anti-correlated modulation with proximity-induced superconductivity. Our results establish twirling moiré superlattices as a versatile platform for nanoscale manipulation of correlated electronic orders in low-dimensional quantum materials.

**Keywords:** Twirling moiré superlattice; Charge density wave; Nanoscale modulation; Strain engineering; Scanning tunneling microscopy

## Introduction

Moiré superlattices formed by stacking two-dimensional van-der-Waals materials with a twist angle have emerged as a powerful platform for engineering electronic structures and exploring emergent quantum phenomena[1–6]. By introducing long-wavelength periodic potentials into uniform atomic lattices, moiré superlattices reconstruct electronic bands into moiré minibands and strongly modify many-body interactions[7–9], enabling a variety of correlated phases, including superconductivity[10,11], correlated insulating states[12–14], and charge ordering[15–17]. Such moiré-band reconstruction has established a central paradigm for manipulating quantum phases in low-dimensional materials. Beyond global band engineering, increasing attention has recently been devoted to local modulation of correlated electronic phases within moiré superlattices[18–25], because spatially varying electronic environments may enable emergent quantum states and functionalities inaccessible in uniform materials. These developments highlight the growing opportunity to manipulate collective quantum states at the nanoscale.

Beyond conventional rigid moiré superstructures, lattice relaxation and strain reconstruction in the small-angle and near-commensurate regime generate reconstructed moiré landscapes with continuously varying local strain, stacking configuration, and interfacial coupling[26–32]. Recent theoretical studies have further revealed that such reconstructed moiré superlattices can spontaneously develop chiral twirling textures through symmetry breaking of the moiré domain-wall network, representing a new class of reconstructed moiré structures with continuously varying local environments[33,34]. Such continuously varying local environments are particularly attractive for correlated electronic phases, because collective electronic orders, such as charge density waves (CDW) and superconductivity, are highly sensitive to such variations[35–39]. Therefore, unlike disorder-driven electronic inhomogeneity in correlated materials, reconstructed moiré superlattices provide a unique platform for deterministic nanoscale manipulation of collective phases within a single moiré unit cell. Despite the recent advances in reconstructed moiré physics[40–45], how such continuously reconstructed moiré landscapes reshape intertwined collective electronic states remains largely unexplored.

Here, we demonstrate an intra-moiré manipulation of CDW states by construction of a twirling superlattice in twisted monolayer (1L)-$VTe_2$/$NbSe_2$ heterostructures grown by molecular beam epitaxy. Owing to the near lattice commensurability and small twist angle between $VTe_2$ and $NbSe_2$, the reconstructed moiré develops twirling textures accompanied by pronounced local strain variation. Using scanning tunneling microscopy and spectroscopy, we show that the intrinsic CDW of monolayer $VTe_2$ undergoes pronounced intra-moiré reconstruction into inequivalent local regions with distinct stability and coherence. In particular, one local region exhibits strongly suppressed CDW order, while another sustains enhanced short-range CDW correlations persisting to room temperature. First-principles calculations further reveal that the reconstructed CDW landscape originates from strong local strain variation within the twirling moiré superlattice, where compressive strain substantially stabilizes the charge order. In addition, the reconstructed CDW landscape exhibits competing interplay with proximity-induced superconductivity in the underlying $NbSe_2$, forming a continuously nanoscale modulation of intertwined collective orders.

## Results and discussion

Monolayer 1T-$VTe_2$ has attracted considerable attention as a prototypical two-dimensional correlated material hosting a commensurate 4×4 CDW below ~190 K[46–52]. Although the microscopic origin of the CDW remains under debate, accumulating experimental and theoretical studies suggest that strong electron–lattice coupling plays a dominant role. Owing to this strong coupling, the CDW is highly sensitive to lattice strain, dimensionality, and substrate interactions, making monolayer $VTe_2$ an ideal platform for investigating how reconstructed moiré environments reshape collective electronic states. Meanwhile, layered 2H-$NbSe_2$ is a prototypical van-der-Waals superconductor with well-established charge ordering and superconductivity. Owing to their closely matched hexagonal symmetry and similar lattice constants, i.e., in-plane lattice constants of ~3.46 Å and ~3.55 Å for $NbSe_2$ and $VTe_2$, respectively (Figure S1), twisting $VTe_2$ and $NbSe_2$ provides an ideal platform for constructing reconstructed moiré superstructures. As schematically illustrated in Figure 1a, monolayer 1T-$VTe_2$ is epitaxially grown on bulk 2H-$NbSe_2$ substrate with a small twist angle. Due to the near lattice commensurability between the two materials, the interfacial mismatch cannot be accommodated by a simple rigid moiré modulation. Instead, substantial lattice relaxation and strain reconstruction develop across the interface, giving rise to continuously varying twirling moiré textures.

We first characterize the structure of the reconstructed moiré superlattice in the 1L-$VTe_2$/$NbSe_2$ heterostructure. LEED pattern shows sharp diffraction spots of $VTe_2$, confirming its high crystallinity (Figure 1b). Large-area STM topography reveals $VTe_2$ islands on the $NbSe_2$ surface with relatively uniform morphology (Figure 1c). The height profile exhibits thickness of approximately 0.7 nm, indicating the monolayer characteristic of the $VTe_2$ adlayer (Figure S2). A zoom-in STM image displays a pronounced distorted moiré modulation with a characteristic periodicity of approximately 9.8 nm (Figure 1d). Unlike conventional rigid moiré superlattices with smoothly varying periodic contrast, the observed moiré pattern exhibits continuously curved features resembling twirling textures (Figure 1e), suggestive of substantial lattice reconstruction within a single moiré unit cell.

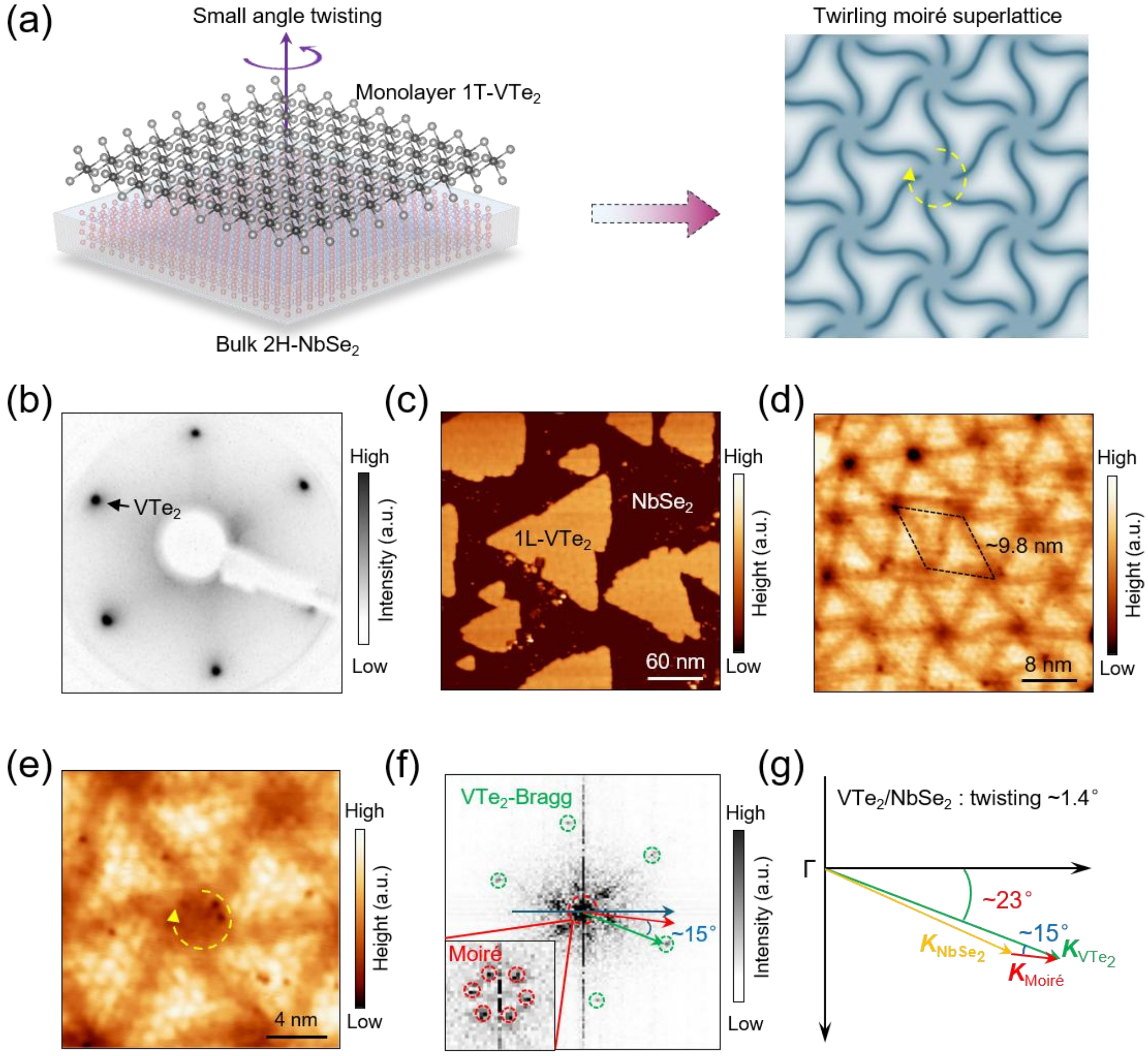

**Figure 1. Construction of twirling moiré superlattices in twisted 1L-$VTe_2$/$NbSe_2$ heterostructure.** (a) Schematic illustration of monolayer 1T-$VTe_2$ epitaxially grown on bulk 2H-$NbSe_2$ with a small twist angle. Lattice reconstruction induces a spatially varying strain field, giving rise to a twirling moiré texture. (b) LEED pattern, showing sharp diffraction

spots of $VTe_2$. (c) Large-scale STM image ($V_s$=-2.0 V, $I_t$=1.0 nA), showing monolayer $VTe_2$ on the surface of $NbSe_2$ substrate. (d,e) Zoom-in STM images of the monolayer $VTe_2$ ($V_s$=-1.5 V, $I_t$=0.05 nA), showing the twirling moiré structure with a large periodicity of ~9.8 nm. (f) The corresponding FFT pattern of (d), showing $VTe_2$ Bragg peaks (green circles) and moiré peaks (red circles, inset). The relative rotation between the $VTe_2$ lattice vector and the moiré modulation is ~15°. (g) Schematic illustration of the reciprocal-space relationship between the three wavevectors of $VTe_2$, $NbSe_2$, and moiré pattern. Based on the vector relation $\boldsymbol{K}_{VTe2} = \boldsymbol{K}_{NbSe2} + \boldsymbol{K}_{moiré}$, the twist angle between $VTe_2$ and $NbSe_2$ is estimated to be ~1.4°.

To understand the moiré geometry, we perform fast Fourier transform (FFT) analysis. In addition to the Bragg peaks of $VTe_2$, low-momentum moiré peaks are also resolved near the center of the FFT pattern (Figure 1f). Notably, the moiré modulation is rotated by approximately 15° relative to the $VTe_2$ lattice vector. Based on the reciprocal-space vector relation $\boldsymbol{K}_{VTe2} = \boldsymbol{K}_{NbSe2} + \boldsymbol{K}_{moiré}$ illustrated in Figure 1g, the twist angle between $VTe_2$ and $NbSe_2$ is estimated to be ~1.4°, placing the system in the small-angle and near-commensurate regime where lattice reconstruction effects could become significant. We further construct a commensurate twisted $VTe_2/NbSe_2$ heterostructure supercell, which generate a moiré periodicity of 9.89 nm at a twist angle of 1.43° (Table S1). The rigid moiré superlattice exhibits relatively smooth periodic modulation, whereas the fully relaxed atomic structure develops characteristic twirling moiré textures (Figure S3), in excellent agreement with the STM observations. Such twirling reconstruction emerge from spontaneous symmetry breaking of the moiré networks, which could give rise to a continuously varying strain and stacking landscape within a single moiré unit cell, providing an ideal platform for investigating how local structural reconstruction reshapes collective electronic states.

We next investigate the impact of the twirling moiré superlattice on the electronic ordered states of monolayer $VTe_2$. Monolayer $VTe_2$ hosts a 4×4 CDW transition below ~190 K driven by strong electron-lattice interactions (Figure S4). Remarkably, the reconstructed moiré landscape spatially modulates this intrinsic CDW, fragmenting it within individual moiré unit cells. Figure 2a shows an atomically resolved STM image acquired at 0.4 K within a representative twirling-textured moiré region. Instead of a uniform long-range 4×4 modulation, the CDW is reconstructed into three inequivalent local regions: two neighboring distorted triangular domains (regions A and B) display pronounced short-range CDW order, whereas the central region H exhibits strongly suppressed CDW intensity.

The corresponding FFT pattern further confirms the short-range nature of the CDW (Figure 2b). In addition to the Bragg peaks of $VTe_2$, diffuse peaks appear around the expected 4×4 CDW wavevectors, indicating spatial fragmentation and absence of long-range coherence. Spatially averaged tunneling spectra acquired in the three regions (Figure 2c) reveal distinct local electronic properties associated with the inequivalent moiré environments. Regions A and B exhibit nearly identical spectral features, whereas region H shows a modified electronic structure with enhanced density of states around +750 mV, indicating that the reconstructed moiré landscape modifies the local electronic states. Moreover, the electronic origin of the fragmented CDW order is further confirmed by the characteristic phase reversal of the modulation around ±100 meV in the energy-resolved d$I$/d$V$ linecut (Figure S5). These observations demonstrate a pronounced intra-moiré reconstruction of the CDW state induced by the twirling superlattice.

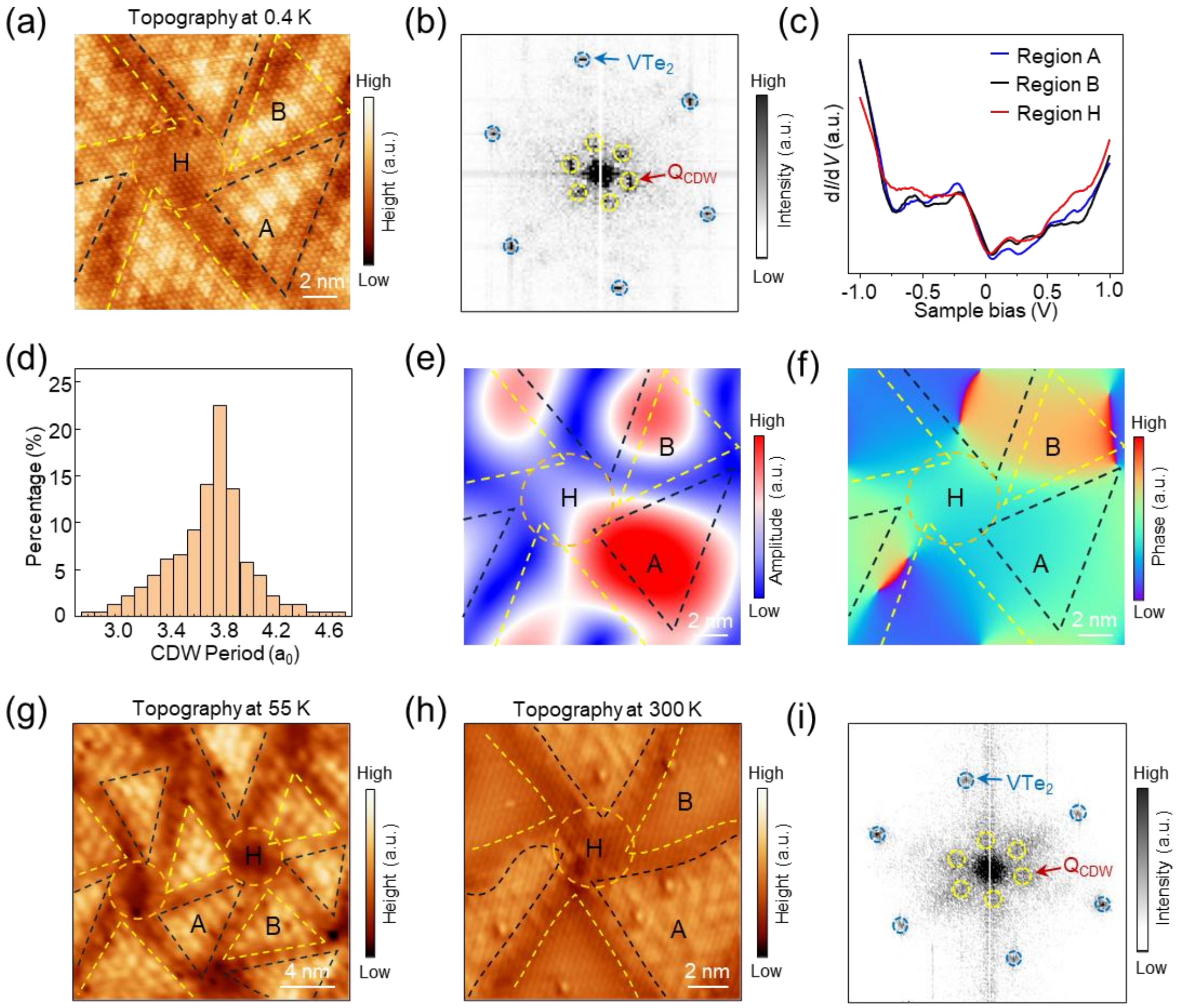

**Figure 2. Modulation of the CDW states in the twirling moiré structure.** (a) Atomically-resolved STM image of a twirling moiré region acquired at 0.4 K ($V_s$=-1.5 V, $I_t$=0.05 nA), showing three inequivalent local regions within a single

moiré unit cell. Regions A and B correspond to neighboring distorted triangular domains exhibiting CDW order, while central region H shows a strong suppression of the CDW. (b) Corresponding FFT pattern of (a), showing Bragg peaks of $VTe_2$ (blue circles) and additional diffuse features associated with short-range CDW orders (yellow circles). (c) The spatially-averaged d$I$/d$V$ spectra ($V_s$=-1.0 V, $I_t$=0.05 nA, $V_{mod}$=5 mV) acquired in the three local regions, Regions A and B exhibit similar spectral features, while region H shows a slightly modified electronic structure with an enhanced density of states around +750 mV. (d) Statistical analysis of the CDW periodicity in regions A and B, showing a quasi-periodicity of about 3.8 times the $VTe_2$ lattice constant $a_0$. (e,f) Spatial distribution of the CDW amplitude (e) and phase (f), showing pronounced amplitude variation and the absence of long-range phase coherence. (g) STM topography acquired at 55 K ($V_s$=-1.0 V, $I_t$=0.05 nA). (h) Atomically-resolved STM image of a twirling moiré region acquired at 300 K ($V_s$=-0.3 V, $I_t$=0.3 nA), showing that short-range CDW order persists only in region A. (i) Corresponding FFT pattern of (h), showing Bragg peaks of $VTe_2$ (blue circles) and diffuse CDW features (yellow circles).

We then quantitatively analyzed the reconstructed CDW modulation by examining the local CDW periodicity in regions A and B. The dominant CDW modulation exhibits a quasi-periodicity of approximately 3.8 times of the pristine $VTe_2$ lattice constant ($a_0$~3.55 Å), deviating from the commensurate 4×4 modulation of the pristine monolayer (Figure 2d). Such deviation suggests that the reconstructed moiré landscape modifies both CDW coherence and the local charge ordering wavevector. Real-space maps of the CDW amplitude and phase, extracted via Fourier filtering, reveal pronounced amplitude variations correlated with the moiré texture and the absence of global phase coherence across the moiré unit cell (Figure 2e,f). These results demonstrate that the long-range CDW state evolves into a fragmented, short-range ordered configuration under the reconstructed moiré potential.

To further examine the local stability of the fragmented CDW state, we performed STM measurements at varying temperatures. While the short-range CDW orders are visible within both regions A and B up to 55 K (Figure 2g), short-range CDW correlations persist to room temperature only within one type of distorted triangular domain (Figure 2h). The corresponding FFT pattern (Figure 2i) still exhibits diffuse CDW features in addition to the Bragg peaks of $VTe_2$, confirming the persistence of short-range charge order at room temperature. The observation of spatially selective CDW survival at elevated temperatures indicates that the reconstructed twirling moiré landscape strongly reshapes the local stability of charge order within a single moiré unit cell, resulting in coexisting electronic regions with substantially different CDW robustness and coherence.

The reconstructed CDW observed here differs fundamentally from the previously reported CDW state in pristine monolayer $VTe_2$. Previous STM studies have established that monolayer $VTe_2$ develops a spatially homogeneous long-range 4×4 CDW below the transition temperature (~190 K), which is generally attributed to strong electron–lattice coupling and is highly sensitive to lattice strain. In contrast, the reconstructed twirling moiré landscape continuously modifies the local structural environment within a single moiré unit cell, transforming the intrinsic long-range CDW into spatially differentiated local phases with distinct coherence, stability, and ordering wavevector. The emergence of locally suppressed CDW regions together with robust short-range CDW correlations persisting up to room temperature demonstrates that the reconstructed moiré landscape not only redistributes the CDW amplitude, but fundamentally reconstructs the local free-energy landscape governing charge ordering.

Given the strong sensitivity of the CDW in monolayer $VTe_2$ to lattice distortion and electron–lattice coupling, we next investigate how the reconstructed twirling moiré landscape modifies the local structural environment. Statistical analysis of the lattice constants extracted from atomically resolved STM images reveals three distinct structural regions within a single moiré unit cell (Figure 3a-3c). The dominant lattice constants are ~3.42 Å and ~3.51 Å in regions A and B, respectively, whereas the central region H exhibits a significantly expanded lattice constant of ~3.65 Å. The local lattice therefore evolves continuously from compressive (region A) to weakly compressive (region B) and finally to tensile (region H), directly demonstrating substantial strain reconstruction within the twirling moiré supercell. Remarkably, this structural evolution coincides with the experimentally observed evolution of the CDW, from robust short-range order in region A, to weakened correlations in region B, and ultimately to complete suppression in region H, suggesting that the reconstructed local strain environment governs the stability of the charge-ordered phase.

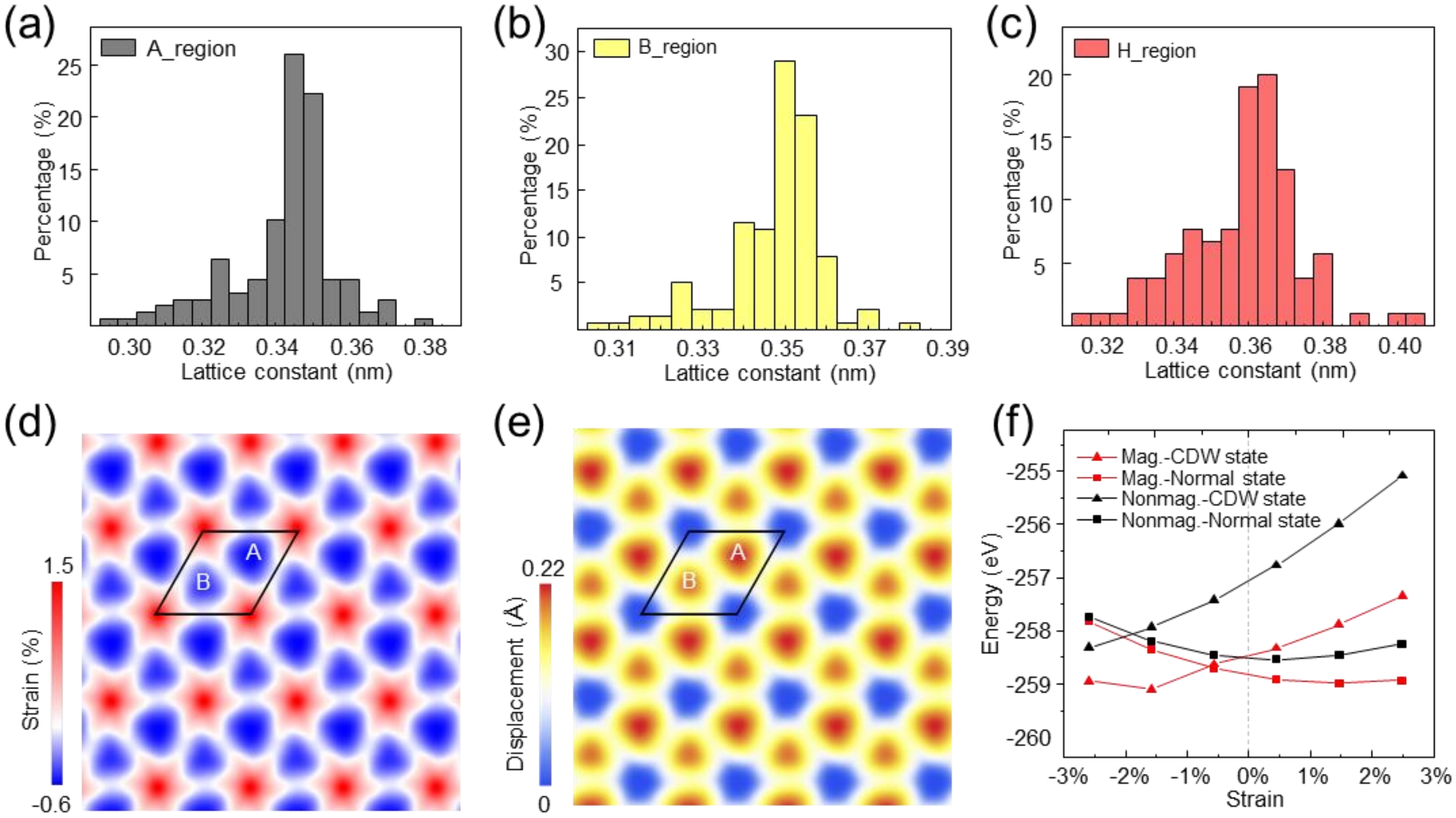

**Figure 3. Local strain analysis in the twisted moiré structure.** (a-c) Statistical analysis of the lattice constant of $VTe_2$ in the three inequivalent local regions, showing dominant lattice constants of ~3.42 Å and ~3.51 Å in regions A and B, respectively, while a significantly larger value of ~3.65 Å in region H. (d) Spatial map of the in-plane strain of the top-layer Te atoms, showing pronounced strain variation across the moiré unit cell. (e) Out-of-plane displacement of the top-layer Te atoms, showing correlated lattice reconstruction accompanying the in-plane strain. (f) Calculated biaxial strain dependence of the relative stability between the CDW and normal phases. Triangles and squares denote the CDW and normal states, respectively, while red and black curves correspond to calculations with and without magnetization. The dashed line marks the reference lattice constant ($a_0$ ~3.55 Å) of $VTe_2$ in this system.

To further visualize the reconstructed structural landscape, we calculated the spatial distribution of the local strain based on the relaxed heterostructure model, focusing on the top-layer Te atoms to facilitate direct comparison with STM measurements. The local in-plane strain is defined as the average distance from each Te atom to its six nearest neighbors relative to the equilibrium lattice constant of $VTe_2$ (~3.55 Å). The resulting strain map (Figure 3d) exhibits pronounced spatial variation across the twirling moiré unit cell, with the central region H experiencing the largest tensile strain while the neighboring triangular domains (regions A and B) remain under compressive strain, in agreement with STM results. Notably, the absence of $C_{2z}$ symmetry in monolayer $VTe_2$ renders the two compressive triangular regions inequivalent, producing asymmetric strain and shape. Furthermore, out-of-plane displacements further reflect this asymmetry (Figure 3e). The central region corresponds to the lowest height, whereas the CDW-hosting regions A and B occupy slightly elevated positions with distinct vertical displacements. This coupled in-plane and out-of-plane lattice reconstruction generates three inequivalent local structural environments that provide the structural basis for the intra-moiré reconstruction of the CDW.

To elucidate the microscopic origin of the reconstructed CDW, first-principles calculations of the relative energies between the CDW and normal phases under biaxial strain were performed (Figure 3f). Since the magnetic ground state of monolayer $VTe_2$ remains controversial, both magnetic and nonmagnetic calculations were performed to comprehensively evaluate the strain dependence of the CDW phase. In both cases, compressive strain substantially stabilizes the CDW phase, whereas tensile strain continuously weakens the energetic stability of the charge-ordered state and progressively favors the normal phase. Notably, calculations including magnetic ordering shift the CDW-normal crossover closer to the equilibrium lattice constant, suggesting that magnetic interactions further promote CDW formation. Although the magnetic configuration is energetically favored in our calculations, both calculation schemes consistently reproduce the same strain-dependent evolution of the CDW, indicating that the experimentally observed behavior is primarily governed by the reconstructed local strain rather than the assumed magnetic state.

The calculated strain dependence agrees remarkably well with the experimental observations. Region A, which experiences the strongest local compression, hosts the most robust short-range CDW correlations that persist up to room temperature. Region B, subjected to weaker compression, exhibits reduced CDW stability, whereas the tensile central region H displays complete suppression of the charge order. This systematic evolution mirrors the calculated monotonic decrease in CDW stability with increasing tensile strain, establishing a direct correspondence between the continuously reconstructed local strain environment and the intra-moiré evolution of the CDW. These results demonstrate that the twirling moiré reconstruction does not merely modulate the CDW amplitude; instead, it continuously tunes the local energetic stability of the charge-ordered phase, transforming an intrinsically homogeneous long-range CDW into spatially differentiated local electronic states within a single moiré unit cell. Although the present work establishes a direct correspondence between the reconstructed local environment and the evolution of the CDW, the reconstructed moiré superlattice possesses only one characteristic domain size set by lattice relaxation. Consequently, the present experiments do not allow us to isolate a possible critical domain size for sustaining the CDW, which remains an interesting question for future investigations using twist-angle-tunable moiré superlattices.

Having established that the reconstructed twirling moiré landscape continuously reshapes the local stability of the CDW through strain reconstruction, we finally examine whether this reconstructed electronic landscape also influences other collective electronic phases. Owing to the superconducting $NbSe_2$ substrate, the $VTe_2/NbSe_2$ heterostructure provides a unique opportunity to investigate how the reconstructed CDW landscape affects the local proximity-induced superconducting response. Figure 4a shows a representative STM topography of the twirling moiré superlattice. Remarkably, spatially resolved tunneling spectroscopy performed across alternating A and B regions reveals pronounced modulation of the superconducting spectral features within the moiré unit cell (Figure 4b,c), indicating that the proximity-induced superconducting state is strongly influenced by the reconstructed local environment.

To visualize the spatial evolution of superconductivity, we extracted the zero-energy conductance $G(\boldsymbol{r},0)$, superconducting coherence peak intensity $G(\boldsymbol{r},\Delta)$, and superconducting gap depth $H(\boldsymbol{r})$, as shown in Figure 4d-4f. All three quantities exhibit pronounced intra-moiré modulation that closely follows the twirling moiré texture. Remarkably, the superconducting response evolves in an anti-correlated manner with the reconstructed CDW landscape: regions hosting stronger CDW correlations exhibit enhanced zero-bias conductance together with suppressed coherence peaks and reduced superconducting gap depth, whereas regions with weakened CDW order display relatively stronger superconducting signatures. This spatial correspondence indicates that the reconstructed moiré landscape simultaneously governs both collective electronic orders.

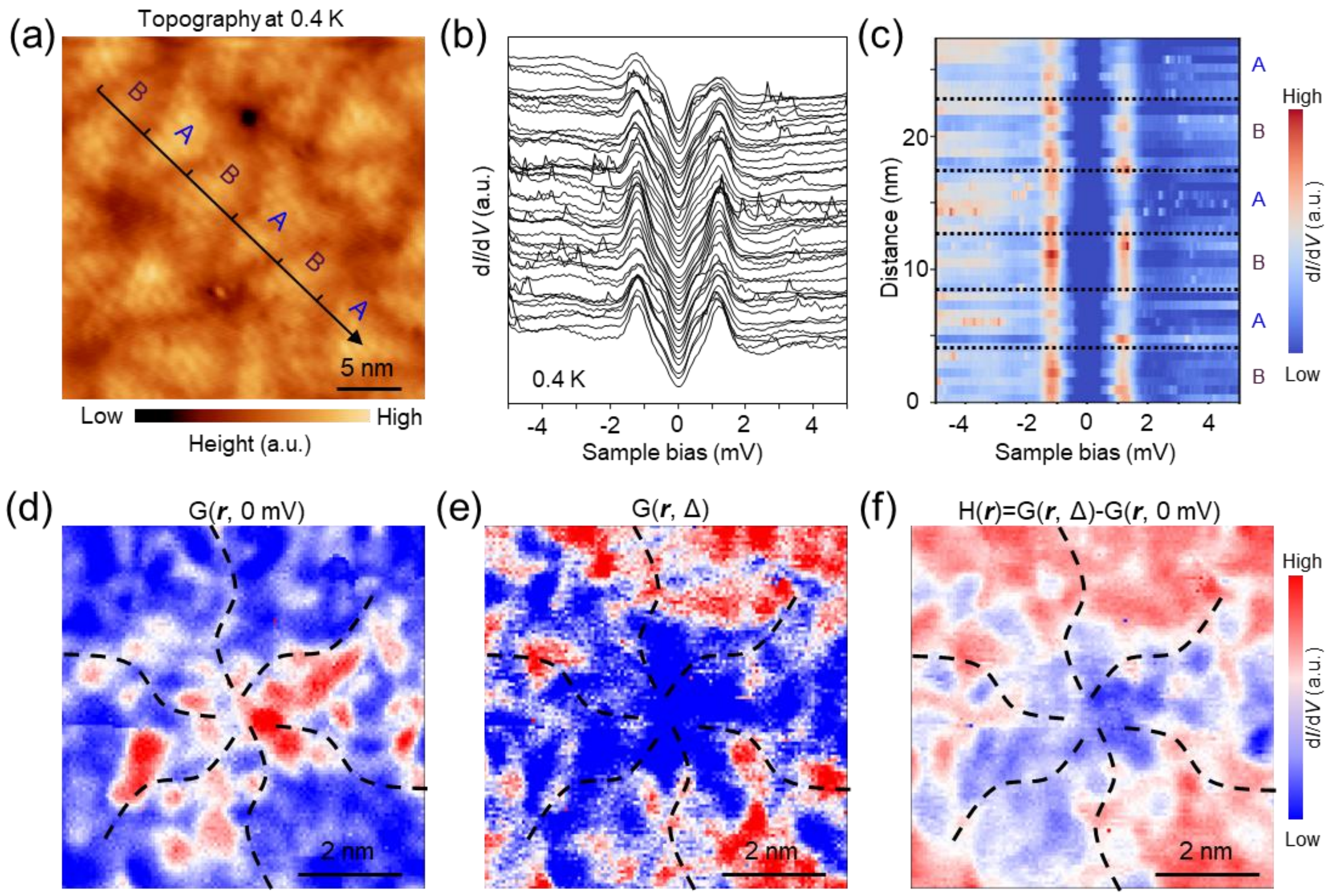

**Figure 4. Influence of the reconstructed CDW on proximity-induced superconductivity.** (a) STM image displaying twirling moiré superlattice ($V_s$=-1.5 V, $I_t$=50 pA). (b-c) The waterfall plot (b) and corresponding intensity map (c) of the d$I$/d$V$ linecut acquired along the line crossing alternating A and B regions indicated in the topography in (a), revealing periodic modulation of the superconducting gap within the moiré unit cell. ($V_s$=-5 mV, $I_t$=1 nA, $V_{mod}$=0.05 mV) (d-f) The d$I$/d$V$ maps at the zero-energy conductance $G(\boldsymbol{r}, 0\ \text{mV})$ (d), coherence peak $G(\boldsymbol{r}, \Delta)$ (e) and superconducting gap depth $H(\boldsymbol{r})$ (f), showing pronounced intra-moiré modulation of the proximity-induced superconducting state. The black dash curves outline the twirling moiré superlattice for eye-guide ($V_s$=-5 mV, $I_t$=1 nA, $V_{mod}$=0.05 mV).

Rather than representing an independent phenomenon, the superconducting modulation provides an additional manifestation of the reconstructed electronic landscape established by the twirling moiré superlattice. As demonstrated above, compressive strain locally stabilizes the CDW phase and enhances short-range charge ordering, which partially suppresses the low-energy electronic states available for superconducting pairing. By contrast, regions with weakened CDW order retain enhanced low-energy spectral weight and therefore support a relatively stronger superconducting proximity effect. The spatial anti-correlation between the reconstructed CDW and superconductivity therefore reflects the local balance between two competing collective electronic orders governed by the same reconstructed structural environment. These results further demonstrate reconstructed twirling moiré superlattices as a promising platform for deterministic nanoscale manipulation of multiple competing quantum states in two-dimensional materials.

## Conclusions

We have realized a twirling moiré superlattice in twisted 1L-$VTe_2$/$NbSe_2$ heterostructures in the small-angle and near-commensurate regime. Distinct from conventional rigid moiré superlattices, the reconstructed moiré landscape exhibits strongly nonuniform local strain and stacking environments. As a result, the intrinsic long-range CDW order of monolayer $VTe_2$ undergoes pronounced intra-moiré reconstruction into inequivalent local phases with distinct stability and coherence within a single moiré unit cell, which we attribute to the strongly varying local strain distribution generated by the twirling moiré superstructure. Furthermore, the reconstructed CDW state exhibits competing interplay with proximity-induced superconductivity from the $NbSe_2$ substrate. Our work establishes reconstructed twirling moiré superlattices as a promising platform for deterministic nanoscale manipulation of collective electronic states beyond conventional moiré band engineering.

## Methods

**Sample preparation.** The monolayer $VTe_2$ was epitaxially grown on freshly-cleaved $NbSe_2$ substrates by molecular beam epitaxy approach under ultra-high vacuum conditions (UHV, base pressure ~$5 \times 10^{-10}$ mbar). High-quality $NbSe_2$ single crystals used as substrates were fabricated by chemical vapor transport approach. The clean $NbSe_2$ substrates were obtained by cleaving in vacuum and subsequently annealing in UHV at 600 K for several hours. The $VTe_2$ films were grown by e-beam evaporation of V (99.9%, *Goodfellow Cambridge Ltd.*) and simultaneous deposition of atomic Te (99.99%, *Sigma-Aldrich*) from a Knudsen cell at a substrate temperature of 490 K. During growth, the Te flux approximately an order of magnitude greater V flux (Te-rich conditions) is used.

**STM/STS and LEED.** STM/STS measurements were performed in an ultrahigh vacuum ($1\times10^{-10}$ mbar) ultra-low temperature condition equipped with 11 T magnetic field. The stable sample temperature can be kept at a base temperature of 0.4 K and 4.2 K, respectively. The electronic temperature is 620 mK at a base temperature of 400 mK. All the scanning parameters (setpoint voltage and current) of the STM topographic images are listed in the captions of the figures. Unless otherwise noted, the differential conductance (d*I*/d*V*) spectra were acquired by a standard lock-in amplifier at a modulation frequency of 973.1 Hz. Tungsten tip was fabricated via electrochemical etching and calibrated on a clean Au(111) surface prepared by repeated cycles of sputtering with argon ions and annealing at 770 K. LEED was employed with a 4-grid detector (*Omicron Spectra LEED*) in the UHV chamber at room temperature.

**Two-dimensional lock-in technique.** To explore the short-range CDW, we employ a two-dimensional lock-in technique to determine the amplitude and phase of the modulations[53]. For any arbitrary real space image:

$$A(\mathbf{r}) = \sum_{\mathbf{Q}} a_{\mathbf{Q}}(\mathbf{r})\mathrm{e}^{-i\mathbf{Q}\cdot\mathbf{r}}$$

where $a_{\mathbf{Q}}(\mathbf{r})$ is the complex amplitude at wavevector Q and position **r**. Wavevector Q can be extracted from the Fourier transform $A(\mathbf{q})$ by shifting it back to the center and multiplying a Gaussian window with a cut-off length σ in q-space. The approximate complex amplitude in real space $A_{\mathbf{Q}}(\mathbf{r})$ can be obtained by inverse Fourier transform as following:

$$A_{\mathbf{Q}}(\mathbf{r}) = F^{-1}[A_{\mathbf{Q}}(\mathbf{q})] = \int \mathrm{d}\mathbf{R} A(\mathbf{R}) \mathrm{e}^{i\mathbf{Q}\cdot\mathbf{R}} \mathrm{e}^{-\frac{(r-\mathbf{R})^2}{2\sigma^2}}$$

Thus, using this technique, the amplitude $|A_{\mathbf{Q}}(\mathbf{r})|$ and spatial phase $\Phi_{\mathbf{Q}}^{A}(\mathbf{r})$ of the modulation at Q can be written as:

$$|A_{\mathbf{Q}}(\mathbf{r})| = \sqrt{\mathrm{Re}\, A_{\mathbf{Q}}(\mathbf{r})^2 + \mathrm{Im}\, A_{\mathbf{Q}}(\mathbf{r})^2}$$

$$\Phi_{\mathbf{Q}}^{A}(\mathbf{r}) = \tan^{-1} \frac{\mathrm{Im}\, A_{\mathbf{Q}}(\mathbf{r})}{\mathrm{Re}\, A_{\mathbf{Q}}(\mathbf{r})}$$

Since the phase calculated by the algorithm may have a $2\pi$ jump at certain pixels, it is necessary to perform unwrap function on the obtained amplitude and phase information. When we select the wave vector Q to perform 2D-lockin, the signal after low-pass filtering can be approximated as:

$$A(\mathrm{r}) \sim e^{-i(\mathrm{q}_0 - \mathrm{Q})\cdot \mathrm{r}} = e^{-i\Delta \mathrm{q}\cdot\mathrm{r}}$$

The phase can be written as:

$$\Phi_Q(r) = \Delta\mathrm{q} \cdot \mathbf{r} + \Phi_{local}(\boldsymbol{r})$$

If $q_0 = Q$, then only the local phase information $\Phi_Q(\boldsymbol{r}) = \Phi_{local}(r)$ remains. However, if $q_0 \neq Q$, In the phase, there will be an additional linear term $\Delta\mathrm{q} \cdot \mathrm{r}$.

**First-principles calculations**

In the lattice construction and molecular dynamics simulation part, we used the Moirestudio software package[54] to build the moiré lattice structure. To perform the atomistic relaxation, this work employs a deep-learning-based interatomic potential. The workflow consists of three stages: dataset construction, potential training/validation, and model deployment. For the dataset construction, we started from the 2×2 AA stacked supercell ($VTe_2/NbSe_2$) as the initial structure. Configuration space was sampled using a 9×9 grid of in-plane displacements along two independent directions, yielding 81 configurations. Each configuration was structurally relaxed using VASP with IVDW = 11[55], a plane-wave cutoff energy of 480 eV, a 6×6×1 k-point mesh, and an ionic relaxation criterion of forces below 0.005 eV/Å. In the next two steps, all models were trained using DeePMD-kit v3 with PyTorch as the backend[56,57].

The biaxial strain calculations were conducted utilizing first-principles methods rooted in Density Functional Theory. We employed the generalized gradient approximation exchange-correlation potentials[58], supplemented by the projector augmented wave method for electron-ion interaction[59], as implemented in the VASP code[60,61]. All self-consistent calculations were performed with a plane-wave cutoff energy of 500 eV. Geometric optimizations were executed without constraints until the force acting on each atom was less than 0.01 eV/Å, and the energy variation per cell was less than $10^{-5}$ eV. In addition, we ensured a vacuum layer exceeding 20 Å to avert interactions between periodic images. Information pertaining to the Brillouin zone k-points grid with a separation of < 0.03 $Å^{-1}$.

**Data availability**

Data measured or analyzed during this study are available from the corresponding author on reasonable request.

**Supporting Information**

The topographic and height profile characterization of the moiré heterostructure, the rigid structural mode and relaxed structural mode based DFT calculations, the d*I*/d*V* spectra across the CDW modulations, detailed structural calibration information.

**Acknowledgements**

This work is supported by grants from the National Natural Science Foundation of China (62488201, 52572188, 92580202, 12204037, 12547158), the National Key Research and Development Projects of China (2022YFA1204100), the CAS Project for Young Scientists in Basic Research (YSBR-053 and YSBR-003). W.J. thank the Beijing Institute of Technology Research Fund Program for Young Scholars.

**Author Contributions:** H.-J.G., H.C., and H.G. designed the experiments. H.G., Q.F., J.Y.W., S.Y.X., and J.Y.H. fabricated the samples. S.H.L. and H.T.Y. synthesized the high-quality $NbSe_2$ bulk crystals. H.C., Y.H.S., and Z.Y.C. performed STM experiments. W.J., J.Y.D. and Y.K.C. did the theoretical calculations. All of the authors participated in analyzing the data, plotting figures, and writing the manuscript.

**Competing Interests:** The authors declare that they have no competing interests.

**TOC Graphic**

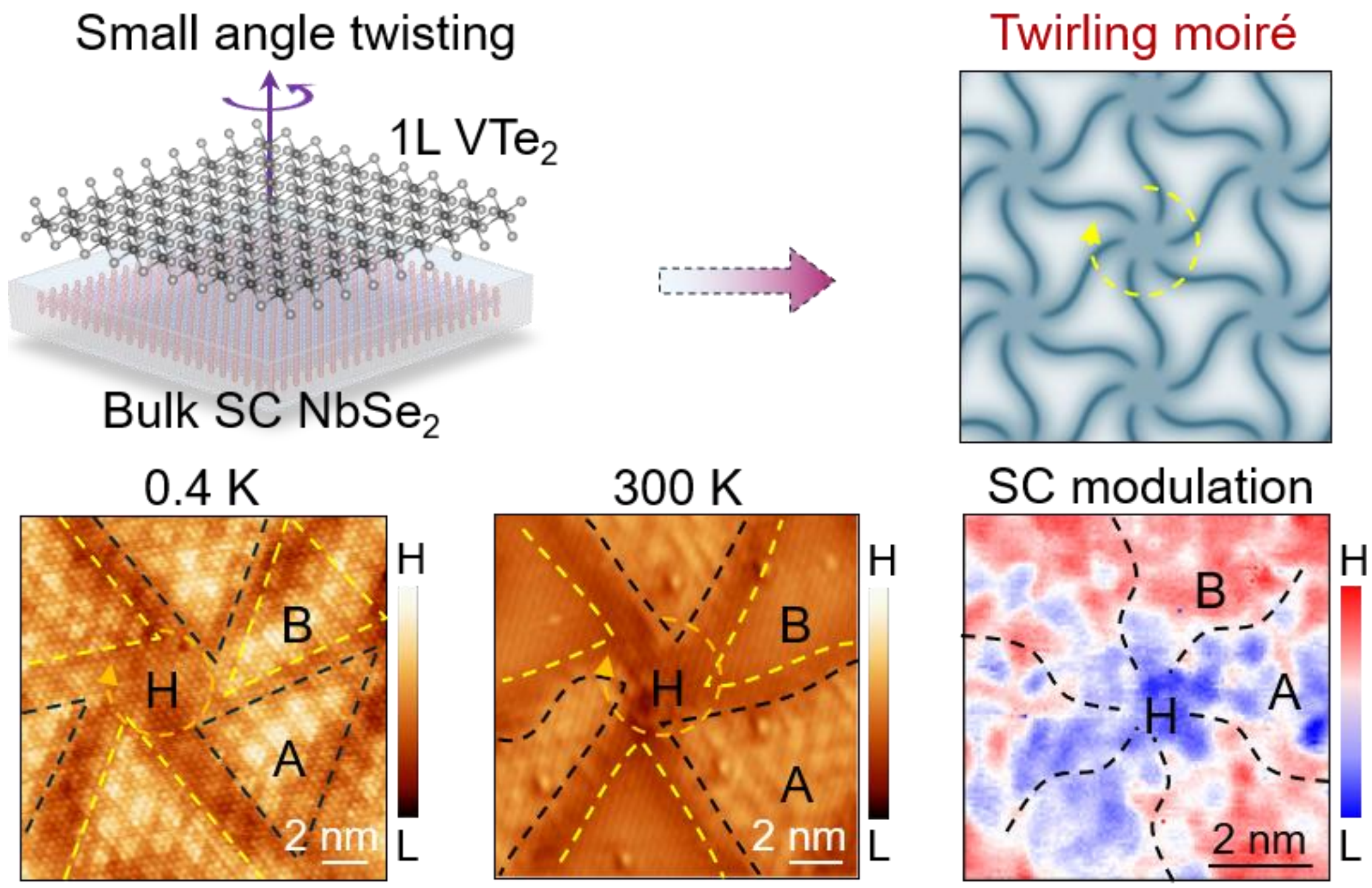

Small angle twisting
1L VTe2
Bulk SC NbSe2
Twirling moiré
0.4 K
300 K
SC modulation
H
B
A
2 nm
L